\documentclass{ecai}
\usepackage{graphicx}
\usepackage{latexsym}

\usepackage{subcaption}
\usepackage{multirow}
\usepackage{makecell}
\usepackage{booktabs}
\usepackage{xcolor}

\usepackage[export]{adjustbox}

\usepackage{tikz}
\usepackage{pgfplots}
\usetikzlibrary{shapes.geometric,shapes.multipart,arrows,positioning,calc,shapes.arrows,pgfplots.statistics}
\pgfplotsset{compat=1.17}

\newcommand{\our}[0]{ProMIL}

\ecaisubmission 

\usepackage{amsmath,amsfonts,amssymb,amsthm,dsfont}
\usepackage[linesnumbered,ruled,vlined]{algorithm2e}
\usepackage{makecell}

\usepackage{amsmath}
\renewcommand{\vec}[1]{\mathbf{#1}}

\usepackage[capitalize,nosort]{cleveref}
\crefname{section}{Sec.}{Secs.}
\Crefname{section}{Section}{Sections}
\Crefname{table}{Table}{Tables}
\crefname{table}{Tab.}{Tabs.}
\Crefname{appsec}{Appendix}{Appendices}
\crefname{appsec}{App.}{Apps.}
\crefname{algocf}{Alg.}{Algs.}
\Crefname{algocf}{Algorithm}{Algorithms}

\def\cost{\mathrm{cost}}

\theoremstyle{remark}

\SetKwInput{KwParam}{Parameter}

\begin{document}

\begin{frontmatter}

\title{ProMIL: Probabilistic Multiple Instance Learning for Medical Imaging}


 \author[AD]{\fnms{\L{}ukasz}~\snm{Struski}%
 	\thanks{Corresponding Author. Email: lukasz.struski@uj.edu.pl.}}
 \author[A]{\fnms{Dawid}~\snm{Rymarczyk}\orcid{}}
 \author[BD]{\fnms{Arkadiusz}~\snm{Lewicki}\orcid{}}
 \author[CD]{\fnms{Robert}~\snm{Sabiniewicz}\orcid{}} 
 \author[AD]{\fnms{Jacek}~\snm{Tabor}\orcid{}}
 \author[A]{\fnms{Bartosz}~\snm{Zieli\'nski}\orcid{}}

 \address[A]{Faculty of Mathematics and Computer Science, Jagiellonian University}
 \address[B]{Faculty of Applied Computer Science, University of Information Technology and Management in Rzeszow}
 \address[C]{Department of Pediatric Cardiology and Congenital Heart Diseases, Medical University of Gdansk}
 \address[D]{UES Ltd.}

\begin{abstract}
Multiple Instance Learning (MIL) is a weakly-supervised problem in which one label is assigned to the whole bag of instances. An important class of MIL models is instance-based, where we first classify instances and then aggregate those predictions to obtain a bag label. The most common MIL model is when we consider a bag as positive if at least one of its instances has a positive label. However, this reasoning does not hold in many real-life scenarios, where the positive bag label is often a consequence of a certain percentage of positive instances. To address this issue, we introduce a dedicated instance-based method called ProMIL, based on deep neural networks and Bernstein polynomial estimation. An important advantage of ProMIL is that it can automatically detect the optimal percentage level for decision-making. We show that ProMIL outperforms standard instance-based MIL in real-world medical applications. We make the code available.
\end{abstract}

\end{frontmatter}

\section{Introduction}

Classification-oriented machine learning typically assumes that each example in a training set has a unique label assigned to it. However, in many practical situations, it is not feasible to label each instance individually. This results in a challenge known as Multiple Instance Learning (MIL)~\cite{dietterich1997solving}, where a bag of instances is associated with a single label, and it is assumed that some instances in the bag are relevant to that label.

MIL represents a powerful approach used in different application fields, mostly to solve problems where instances are naturally arranged in sets or to leverage weakly annotated data~\cite{carbonneau2018multiple}. Its numerous applications include molecule classification task~\cite{dietterich1997solving}, predicting gene functions~\cite{eksi2013systematically}, content-based image retrieval~\cite{vijayanarasimhan2008keywords}, object location~\cite{hoffman2015detector} and segmentation~\cite{muller2012multi}, computer-aided diagnosis and detection~\cite{borowa2022identifying}, document classification~\cite{harris1954distributional}, and web mining~\cite{zhou2005multi}.

Multitude applications of MIL contribute to introducing many new methods, described in the surveys~\cite{carbonneau2018multiple,zhou2004multi}. They generally divide into two types, instance- and representation-based. The former classifies instances and then aggregates those predictions to obtain a bag label. The latter aggregates instance representations and predicts a bag label directly. Therefore, instance-based approaches are more interpretable but usually less accurate than representation-based methods. Most methods are representation-based and work only for the standard assumption, where a bag of instances is considered positive if it contains at least one positive instance. Only a few consider other assumptions described in~\cite{foulds2010review}, like presence-, threshold-, or count-based assumptions.

\begin{figure}[t]
    \centering
    \includegraphics[width=.35\textwidth]{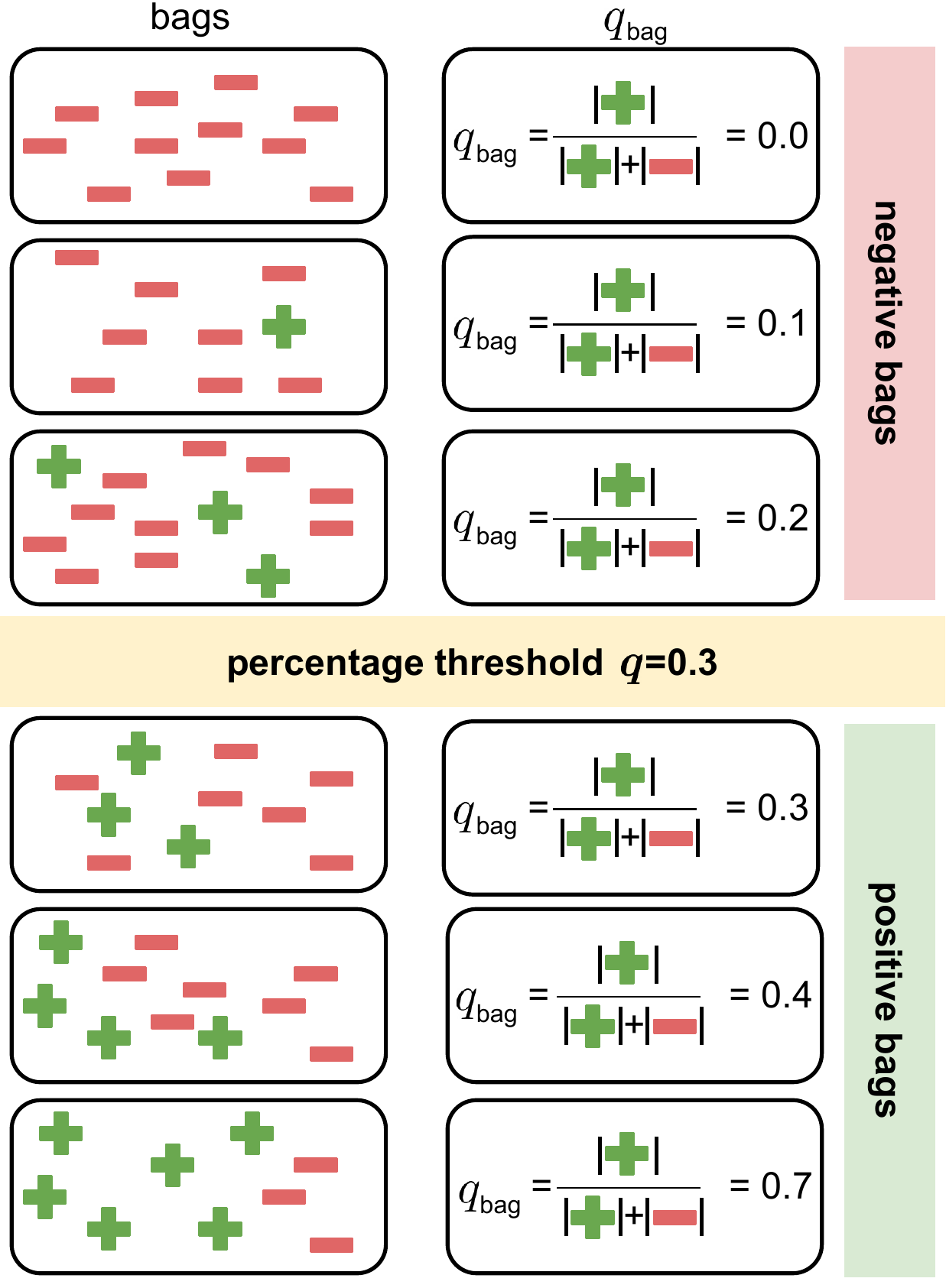}
    \caption{In the percentage-based MIL assumption, each bag has a label indicating whether it is positive or negative based. It is based on the percentage of positive instances, but the information about the percentage threshold and the instance labels are unavailable during training. This example includes six various-sized bags containing positive and negative instances (green pluses and red minuses, respectively), and bags with a percentage of positive instances over threshold $0.3$ are positive. In this paper, we define this assumption and introduce a dedicated method to solve it. }
    \label{fig:teaser}
\end{figure}

\begin{figure*}[t]
    \centering
    \includegraphics[width=0.9\textwidth]{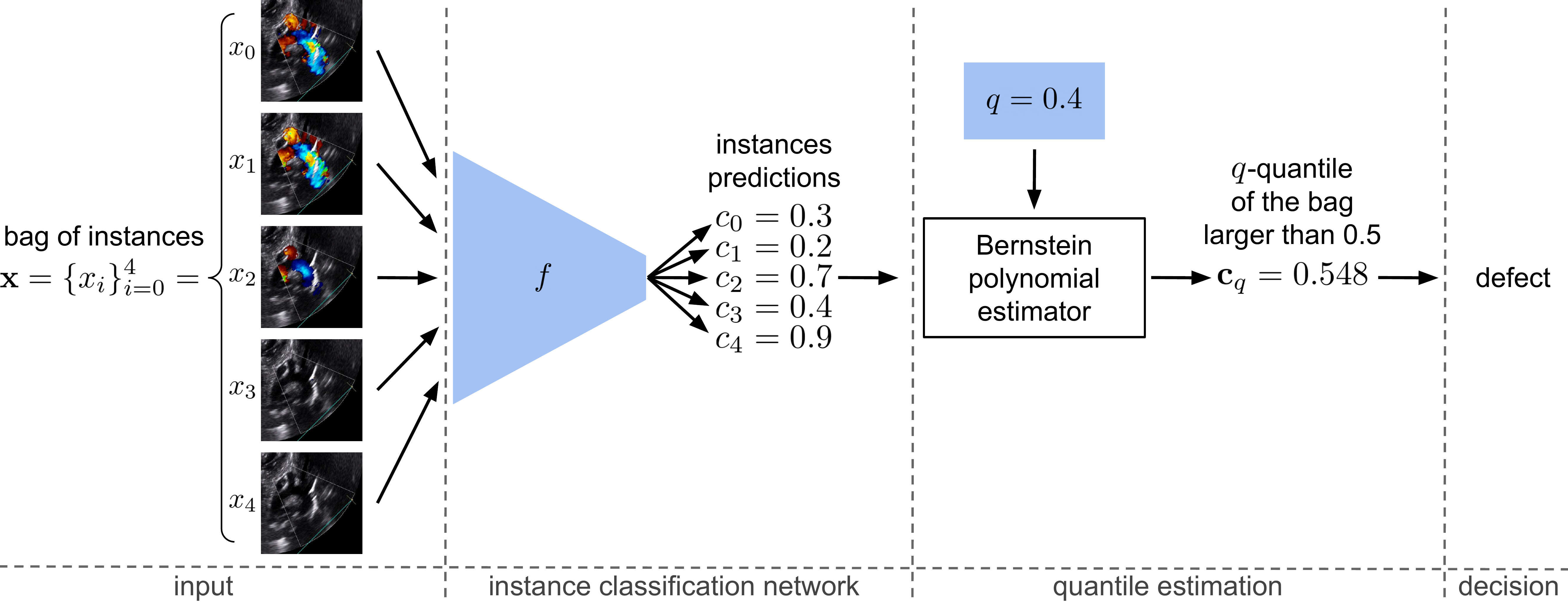}
    \caption{ProMIL model architecture consists of two trainable parameters (here marked as blue boxes), an instance classification network $f$ and parameters $q$. The model takes a bag of instances (e.g. video frames) and passes them through network $f$, obtaining instances predictions $c_0, c_1, \ldots, c_4$. These predictions are then used with the trainable parameter $q$ in the differentiable estimation of the $q$-th quantile ($\vec{c}_q$). If the quantile exceeds $0.5$, then we conclude that the bag is positive (e.g. correspond to a heart defect).}
    \label{fig:ProMIL}
\end{figure*}

In this paper, we consider a specific type of MIL assumption, which we have named the percentage-based assumption. It is motivated by real-world medical problems, such as predicting neutrophils infiltration level in the Geboes scoring system on biopsy images~\cite{geboes2000reproducible}, identifying bacteria strains on microscopic images~\cite{borowa2022identifying}, recording myocardial ischemia on ECG~\cite{sun2012ecg}, or detecting congenital heart defects in newborns on ultrasound videos~\cite{komatsu2021detection}. In percentage-based assumption, a bag is positive if a certain percentage of its instances is positive (however, this threshold is unknown during training). For instance, consider the evaluation of digestive tract health using the NHI scoring system~\cite{li2019simplified}, where a biopsy is given a score of $2$ if more than 50\% of crypts are infiltrated with neutrophils and there is no damage or ulceration in the epithelium.

Besides providing a novel MIL assumption, we introduce a dedicated method \our{} based on the Bernstein polynomial estimation that can be applied to any deep architecture. It classifies each instance separately and analyzes whether the appropriate percentage of those predictions corresponds to a positive label (is above the specific threshold). Importantly, \our{} determines this threshold automatically during training.

We compared our approach with the baseline methods for artificial and real-world datasets, including our database with ultrasound videos of congenital heart defects in newborns that are among the most common developmental abnormalities and often lead to serious health consequences, including increased mortality and complications later in life. Finally, we particularly focus on the interpretable aspect of our method, bearing in mind its potential medical applications.

Our contributions can be summarized as follows:
\begin{itemize}
    \item We define a novel percentage-based MIL assumption motivated by real-work medical problems.
    \item We introduce a method dedicated to this assumption based on the Bernstein polynomial estimator, which automatically discovers the threshold and can be applied to any deep architecture.
    \item Our method overpasses existing instance-based solutions, obtaining high performance while maintaining well-interpretable instance-level predictions.
\end{itemize}

\section{Related Works}

\paragraph{Instance-based MIL.}
Initial instance-based MIL methods trained a classifier by assigning pseudo labels (typically a bag label) to each instance. This approach was introduced in~\cite{maron1997framework} and has been widely applied, e.g. in microscopy image classification~\cite{kraus2016classifying} and drug activity prediction~\cite{zhao2013drug}. However, it can result in noisy labels. Therefore, recent studies focus on enhancing instance-based classifiers with dedicated MIL residual connections~\cite{wang2018revisiting} or dynamic pooling \cite{yan2018deep}.
Other approaches choose only crucial instances for training. Crucial instances can be selected based on classifier outputs~\cite{campanella2019clinical}, centered loss and centroids~\cite{chikontwe2020multiple}, or cluster conditioned distribution~\cite{qu2022dgmil}.
ProMIL is an instance-based MIL method, which, in contrast to existing solutions, is dedicated to percentage-based assumption and can automatically estimate the required threshold.

\paragraph{Representation-based MIL.}
Representation-based methods differ from instance-based approaches because they extract instance-level representations, aggregate them to a bag-level embedding, and use it to classify a bag directly. Most methods use the attention mechanism to assign importance to each instance during representation aggregation. AbMILP~\cite{ilse2018attention}, introducing attention-based pooling, is one of the most popular approaches. It was extended by many mechanisms, including self-attention to detect intra-bag dependencies between instances~\cite{rymarczyk2021kernel}, adversarial training to improve robustness~\cite{hashimoto2020multi}, cluster assignments~\cite{zhu2017wsisa}, siamese networks~\cite{yao2020whole}, attention loss~\cite{shi2020loss}, feature distillation~\cite{zhang2022dtfd}, pyramidal fusion~\cite{li2021dual}, prototypical parts for global interpretability~\cite{rymarczyk2023protomil,yu2023prototypical}, multi-attention~\cite{konstantinov2022multi}, and clustering constrained attention~\cite{lu2021data}.
While there have been attempts to classify single instances while obtaining the bag-level embedding~\cite{myronenko2021accounting}, the predictions obtained through this approach are still noisy. Hence, from the interpretability point of view, instance-based approaches, such as ProMIL, are preferable.

\section{Preliminaries}

\paragraph{Multiple Instance Learning (MIL).}
Multiple Instance Learning (MIL) is a weakly-supervised problem in which one label $\vec{y} \in \mathbb{R}$ is assigned to a bag of instances $\vec{x}=\{x_i\}_{i=0}^n$, where $n$ varies between bags~\cite{foulds2010review}. Moreover, in the \emph{standard MIL assumption}, the label of bag $\vec{y} \in \{0,1\}$, each instance $x_i$ has a hidden binary label $y_i \in \{0,1\}$ (unknown during training), and a bag is positive if at least one of its instances is positive
\begin{equation}
\centering \vec{y} = \begin{cases}
0, & \text{ iff }\sum \limits_{i=0}^{n}y_i = 0, \\ 
1, & \text{ otherwise.} 
\end{cases}
\end{equation}

\paragraph{Bernstein polynomial estimator.}
One of the main advantages of our approach is the ability to automatically discover the threshold in the percentage-based assumption. For this purpose, we use Bernstein polynomial estimator~\cite{cheng1995bernstein,leblanc2012estimating,zielinski2004optimal}. Assuming that we have a set $\vec{p}$ with sorted elements $p_0 \leq p_1 \leq \ldots \leq p_n$, its $q$-quantile estimator is calculated as
\begin{equation}\label{eq:B_estimator_quantile}
  \vec{p}_q = \sum_{k=0}^n \binom{n}{k} q^{n-k}(1-q)^{k} \cdot p_k,
\end{equation}
where $\binom{n}{k}$ is a binomial coefficient.

Compared to the competitive solutions~\cite{kendall1946advanced,rosenblatt1956remarks,van2000asymptotic}, Bernstein polynomial estimator provides more accurate estimates of quantiles for datasets with complex or non-uniform distributions and is computationally efficient, even when implemented using standard numerical methods.

\section{ProMIL}

Probabilistic Multiple Instance Learning (ProMIL) was created with the percentage-based assumption in mind that occurs in many real-world biomedical applications. It assumes that a positive bag is associated with a certain percentage of abnormal instances. Therefore, we classify those instances and then calculate the $q$-quantile of their predictions (in the range $[0,1]$). The bag is considered abnormal if the $q$-quantile is larger than $0.5$ and normal otherwise. From this perspective, a method for quantile estimation should be differentiable to train the model and the optimal value of additional parameter $q$. Therefore, we use Bernstein polynomial estimation~\cref{eq:B_estimator_quantile}.

To formalize our approach, let us assume that $\vec{x}=\{x_i\}_{i=0}^n$ is a bag of instances from the training set with the corresponding label $\vec{y}$. Moreover, let $f$ be the classification network run for each instance $x_i$ separately (see~\Cref{fig:ProMIL}).

During training, presented in~\Cref{alg:method_train}, we take random bag $\vec{x}$ from a training set and pass its instances through network $f$. As a result, we obtain a set of predictions $\vec{c}=\{c_0, c_1, \ldots, c_n\}$. We sort this set so that $c_0 \leq c_1 \leq \ldots \leq c_n$, and then we compute their $q$- and $1-q$-quantile ($\vec{c}_q$ and $\vec{c}_{1-q}$) using~\cref{eq:B_estimator_quantile}. Because $c_i \in [0, 1]$ then, according to properties of Bernstein polynomials~\cite{lorentz1953university} $\vec{c}_q \in [0, 1]$.
Finally, we apply BCE loss to train our model
\begin{equation}\label{eq:cost}
  \cost(\vec{x},\vec{y})= -\vec{y} \log \vec{c}_q - (1-\vec{y}) \log \vec{c}_{1-q},
\end{equation}
where $q$ is a trainable parameter.

\begin{algorithm}[t]
\caption{ProMIL training}\label{alg:method_train}
\KwData{$\vec{X}$, $\vec{Y}$ -- bags from training set and their labels}
\KwParam{$f$ -- instance classification network; $q$ -- quantile (threshold)}
\KwResult{trained $f$ and $q$}
 
 randomly initialize $q$\;
 \While{no converngence}{
  take random bag $\vec{x}=\{x_0, x_1, \ldots, x_n\}$ from $\vec{X}$\;
  take bag label $\vec{y}$ from $\vec{Y}$\;
  calculate $\vec{c}=\{c_i=f(x_i)\}_{i=0}^n$\;
  sort $\vec{c}$ in ascending order\;
  calculate $\vec{c}_q$ ($q$-quantile of $\vec{c}$) using~\cref{eq:B_estimator_quantile}\;
  calculate $\mathrm{cost}$ by applying $\vec{c}_q$ and $\vec{y}$ to~\cref{eq:cost}\;
  update $f$ and $q$ using $\mathrm{cost}$ back-propagation
 }
\end{algorithm}

Most of the calculations described in this algorithm are performed on a logarithmic scale for numerical stability, including \textit{logsumexp} and the binomial coefficient calculated as
$$
{n \choose k} = \exp(\log\Gamma(n+1)-\log\Gamma(k+1)-\log\Gamma(n-k+1)),
$$
where $\log \Gamma$ denotes the natural logarithm of the gamma function. Both \textit{logsumexp} and \textit{lgamma} functions are available in the PyTorch\footnote{\url{https://pytorch.org}} framework.


\section{Experimental Setup}

\subsection{Datasets}

\begin{figure}[t]
    \centering    \includegraphics[width=0.45\textwidth]{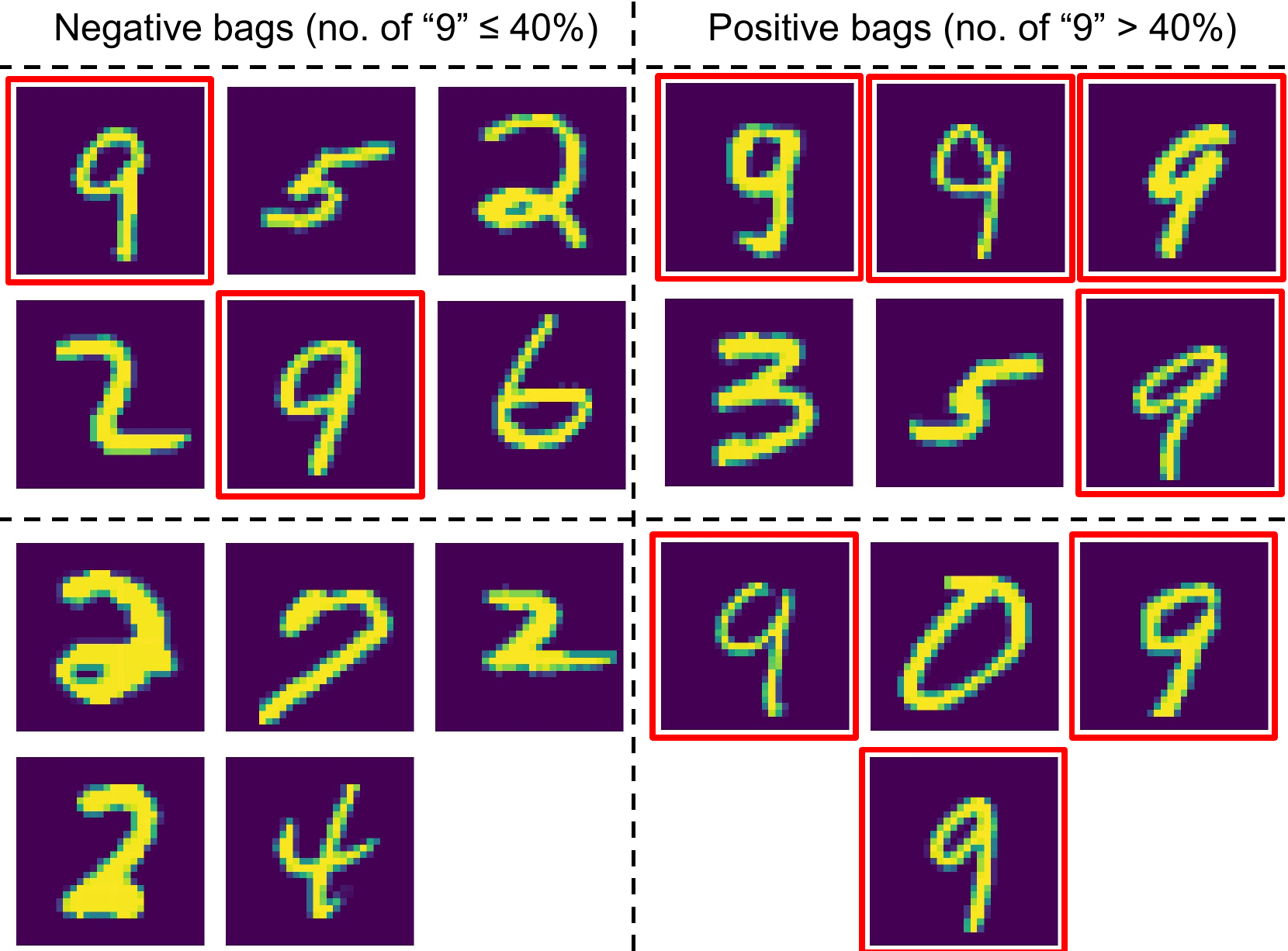}
    \caption{Four examples from the MNIST-bag dataset where the number of instances within a bag is sampled from $\mathcal{N}(5,\,2)$, and the quantile $q=0.4$. In red boxes, we highlight $9$'s corresponding to the positive instances.}
    \label{fig:mil_ds}
\end{figure}

\paragraph{MNIST-bag}
To evaluate the effectiveness of our ProMIL approach for percentage-based MIL assumption, we use the artificial MIL dataset MNIST-bag. The dataset consists of $1000$ bags, each containing instances randomly sampled from a normal distribution with mean of $30$ instances with variance equals to $5$. The percantage of numbers of interests within a bag is taken randomly from a uniform distribution to ensure the balance within a dataset. We permute the order of instances within a bag to maintain permutation invariance. The examples of created bags from the dataset are presented in~\Cref{fig:mil_ds}. 

The dataset is an enhancement of the dataset introduced in~\cite{ilse2018attention} with a different definition of a positive bag to fulfill the percentage-based assumption. We define the bag as positive if at least $q$ fraction of instances from the bag are $9$s. We maintain MNIST division into training and test sets.

\paragraph{Histopathology datasets}
Here we present details on three standard MIL datasets that represent MIL problems in real life, classification of histopathology images: Colon Cancer~\cite{sirinukunwattana2016locality}, Camelyon16~\cite{camelyon16paper}, and TCGA-NSCLC~\cite{bakr2018radiogenomic}. They consist of 100, 399, and 956 H\&E stained whole slide images, respectively. For Colon Cancer, we use patches of the size $32 \times 32$ while for other datasets we use $224 \times 224$. Colon Cancer has $100$ bags with $2,244$ patches, Camelyon has $399$ bags with a mean of $8,871$ patches, and TCGA-NSCLC has $1,016$ bags with a mean of $3,961$ patches. From a medical perspective, the Colon Cancer dataset focuses on detecting epithelial nuclei which are often symptoms of developing disease, Camelyon16 is about the detection of micro- and macro-metastases in lymph nodes which is a prognostic factor for breast cancer growth. While the TCGA-NSCLC dataset is the identification of a cancer subtype which conditions which treatment should be applied. Each of those tasks is of high relevance from a clinical perspective.

\paragraph{Doppler Ultrasound database}
Our proposed approach holds significant potential in analyzing ultrasound videos (USG), which is currently one of the primary diagnostic tools for detecting congenital heart defects in newborns. These defects are among the most common developmental abnormalities and can lead to serious health consequences, including increased mortality and complications later in life.

\begin{figure}[t]
  \centering  \includegraphics[width=0.97\columnwidth]{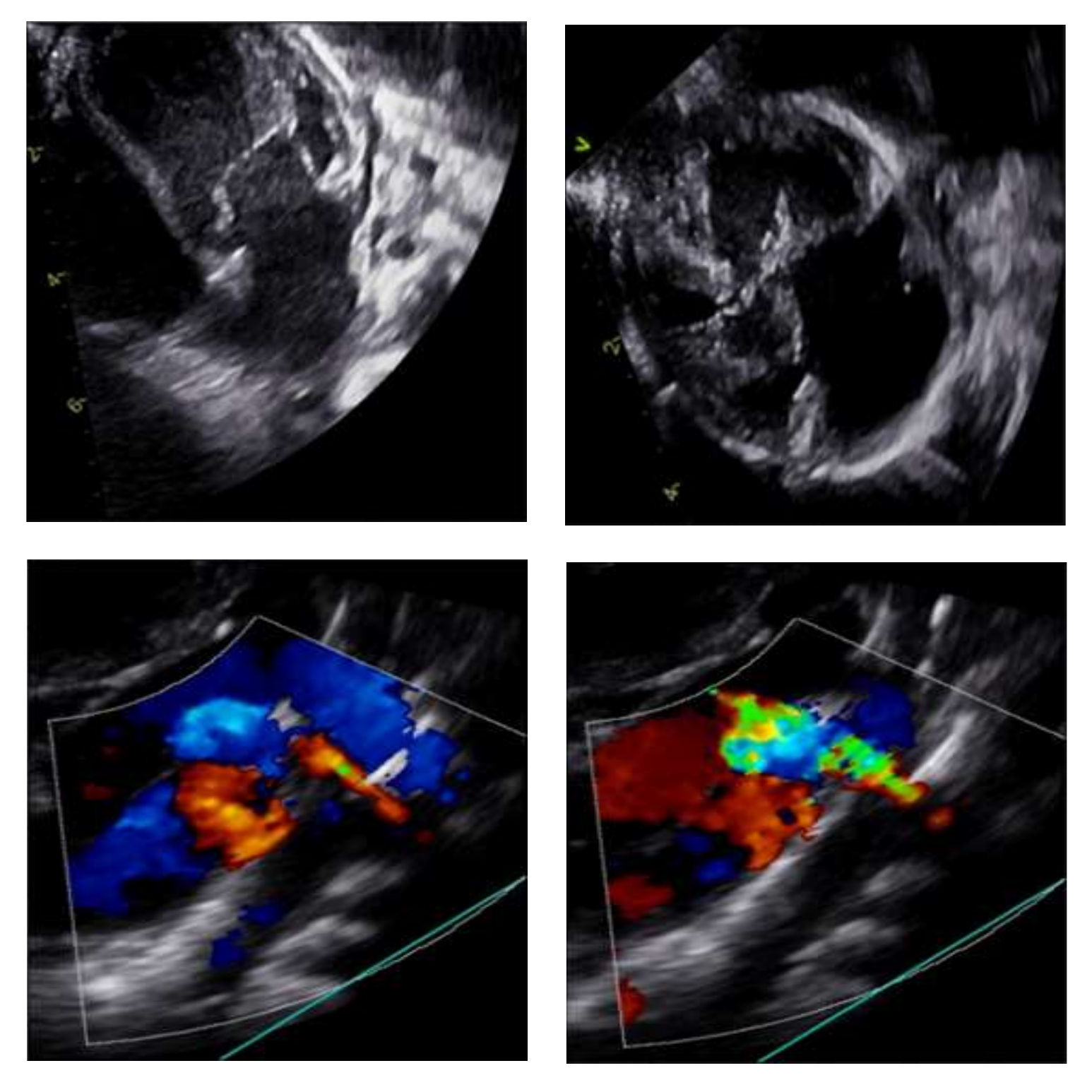}
  \caption{Ultrasound video frames corresponding to congenital heart defects in newborns, such as structural abnormalities and defects in large blood vessels.}
  \label{fig:ultrasond}
\end{figure}

The traditional approach to analyzing USG images involves individual examination and evaluation by qualified physicians specializing in pediatric cardiology. However, this method is time-consuming, requires highly specialized knowledge, and is susceptible to human errors. It is especially problematic for challenging cases, such as Tetralogy of Fallot, hypoplastic left heart syndrome, or defects associated with the pulmonary vein (see~\Cref{fig:ultrasond}).

We performed the Doppler ultrasound on over 250 patients to record blood flow in the heart and detect abnormalities. For some patients, the study was repeated, resulting in a dataset with over 1000 films of varying lengths (ranging from 10 frames to over 240 frames), with over 40\% of the films being without defects. Heart defects specialists labeled each film. Before training the network, the data were randomly split into training and test sets while maintaining class proportions and ensuring that studies from the same patient were not included in both sets.

The training set constituted 85\% of the entire dataset, while the remaining 15\% was reserved for testing. Each frame was scaled to $200\times200$ pixels during network training, and a center crop of 172 pixels was used. During an evaluation, the frames were scaled to $172\times172$ pixels.

\begin{figure*}[t]
    \centering
    \includegraphics[width=0.85\textwidth]{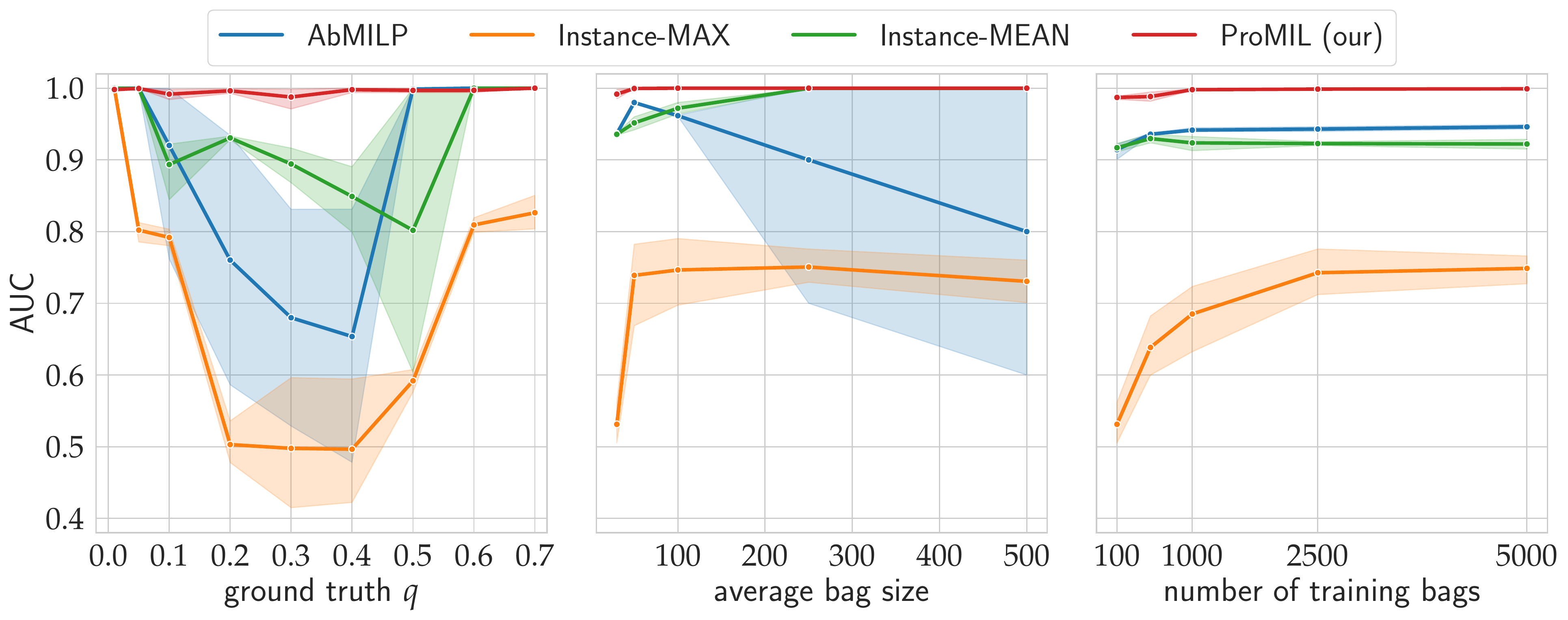}
    \caption{The effectiveness of various MIL methods - AbMILP, Instance-MAX, Instance-MEAN, and ProMIL (our proposed method) on the MNIST-bag dataset was compared. The performance of these methods was analyzed under different scenarios, and the results were plotted. The left image shows the effect of the percentage of positive instances within a bag on the performance of the methods. The middle plot depicts how the bag size influences the effectiveness of MIL methods. The right plot shows the impact of the training dataset size on the model's performance. It can be observed that our proposed ProMIL method consistently outperformed other methods in all scenarios. Moreover, AbMILP was found to be unstable due to overfitting when the number of instances within a bag increased. Lastly, the other methods showed higher variance in the results as compared to our ProMIL method.}
    \label{fig:mnist_results}
\end{figure*}

\subsection{Implementation details}

For the MNIST-bag dataset, we use LeNet-5~\cite{lecun1989handwritten} network to encode the images. It is learned for 100 epochs with early stopping after 15 epochs without a change on a validation set. Training for each scenario was repeated five times. We use Adam optimizer~\cite{kingma2014adam} with learning rate $10^{-4}$ with $\beta_1= 0.99, \beta_2=0.999$, weight decay equals $10^{-5}$, and batch size 1. AbMILP used as a baseline method also uses LeNet5 backbone, consisting of one attention head with embedding size equal to $500$ and classifier hidden space $128$. We use a binary cross entropy as a loss function. 

For Colon Cancer dataset we use backbone from~\cite{sirinukunwattana2016locality}, Adam optimizer with learning rate $10^{-3}$ with $\beta_1= 0.99, \beta_2=0.999$, weight decay equals $10^{-3}$, and batch size $1$. We train a model for $200$ epochs with an early stopping window equal to $25$ epochs. We split the data into ten folds and repeat the training 5 times, similarly as it is done in~\cite{ilse2018attention}. 

When it comes to Camelyon16 and TCGA-NSCLC datasets, we use pretrained in a self-supervised way ResNet18~\cite{ciga2020self} as a backbone and train a model for $200$ epochs with an early stopping window equal to $40$ epochs. Also, we use Adam optimizer with learning rate $5 \cdot 10^{-4}$ with $\beta_1= 0.99, \beta_2=0.999$, weight decay equals $10^{-3}$, and batch size $1$.

For all the aforementioned datasets, we performed a hyperparameter random grid search with the following sets of parameters: learning rate in the range $[10^{-6}, 10^{-2}]$, weight decay in range $[0, 1]$, $q$ in the range $[0.1, 0.5]$.

To compute frame-based predictions for ultrasound videos, we utilized ResNet-18~\cite{he2016deep}, which was pretrained on the ImageNet~\cite{deng2009imagenet} dataset. We replaced the last layer with two fully-connected layers, each having dimensions of $128$ and $64$, respectively. We employed grid search techniques to fine-tune hyperparameters, such as the learning rate (with possible values of $10^{-6}, 10^{-5}, 10^{-4}, 10^{-3}$), dropout rate (searched over values of 0.1, 0.2, 0.3, 0.4), and L2 regularization (with possible values of $0, 10^{-6}, 10^{-5}, 10^{-4}$). Moreover, we experimented with different learning rate schedulers, including CyclicLR, LambdaLR, and without any scheduler.

We trained the models in two stages. In the first stage, we froze the feature extraction part of the pre-trained ResNet-18 model and only trained the two additional linear layers for 25 epochs. In the second stage, we fine-tuned the entire network for 100 epochs. To train the models, we used the Adam optimizer~\cite{kingma2014adam} with default values, similar to other datasets. To accommodate the varying lengths of videos in our dataset, we utilized a batch size of 1.

\section{Results}

\begin{table}
\caption{Results for Colon Cancer dataset, where \textsc{ProMIL} outperforms the baseline methods (\textsc{instance+max} and \textsc{instance+mean}). However, it achieves worse results compared to more advanced techniques that employ attention mechanisms such as \textsc{SA-AbMILP*}. Notice that values for comparison are indicated with ``*'' from~\cite{rymarczyk2023protomil}.}
\centering
\scriptsize 
\begin{tabular}{ccc}
\cmidrule{2-3}
& \multicolumn{2}{c}{\textsc{Colon Cancer}} \\ [0.5ex]
\toprule
\textsc{Method} & \textsc{Accuracy} & \textsc{AUC}  \\ [0.5ex]
\midrule
\textsc{embedding+max*} & $82.4\%\pm 1.5\%$ & $0.918\pm 0.010$ \\ [1ex]
\textsc{embedding+mean*} & $86.0\%\pm 1.4\%$ & $0.940\pm 0.010$\\ [1ex]
\textsc{AbMILP*}  & $88.4\%\pm 1.4\%$ & $0.973\pm 0.007$ \\ [1ex]
\textsc{SA-AbMILP*} & \textit{90.8\%} $\pm$ \textit{1.3\%} & \textit{0.981} $\pm$ \textit{0.007}\\ [1ex]
\textsc{ProtoMIL*} & $81.3\%\pm 1.9\%$ & $0.932\pm 0.014$ \\ [1ex]
\midrule
\textsc{instance+max*} & $84.2\%\pm 2.1\%$ & $0.914\pm 0.010$  \\ [1ex]
\textsc{instance+mean*} & $77.2\%\pm 1.2\%$ & $0.866\pm 0.008$ \\ [1ex]
\textsc{ProMIL (our)} & $\textbf{88.1}\%\pm \textbf{1.7\%}$ & $\textbf{0.969}\pm \textbf{0.010}$\\[1ex]
\bottomrule
\end{tabular}
\label{table:colon}
\end{table}

This section provides the results of experiments conducted on five different datasets. We begin with MNIST-gab, which is a toy dataset that enables a more thorough understanding and analysis of our ProMIL method. Next, we present our findings on three histopathological datasets, which are commonly used benchmarks for MIL algorithms, demonstrating the versatility of our approach. Finally, we introduce a novel heart ultrasound dataset, where we demonstrate the suitability of our ProMIL method for analyzing medical videos. We also include a detailed interpretability analysis of the method conducted with specialized medical doctors to confirm that our models accurately recognize important features.

\begin{table}
\caption{Our ProMIL approach achieves comparable or better results than baseline approaches, specifically \textsc{instance+max} and \textsc{instance+mean}, for the Camelyon16 and TCGA-NSCLC datasets. However, it falls short of matching the performance of more complex models such as TransMIL, which are based on transformer architectures. It is worth noting that ProMIL matches the performance of the \textsc{instance+max} approach. This may be related to the characterization of lymph metastasis (Camelyon16 dataset), which typically covers a small tissue portion. As a result, the ProMIL learns to estimate a small value for $q$ and achieves similar results to the \textsc{instance+max} approach. Notice that values for comparison marked with ``*'' are taken from~\cite{rymarczyk2023protomil}.}
\centering
\scriptsize 
\begin{tabular}{ccccc}
\cmidrule{2-5}
& \multicolumn{2}{c}{\textsc{Camelyon16}} & \multicolumn{2}{c}{\textsc{TCGA-NSCLC}} \\ [0.5ex]
\toprule
\textsc{Method} & \textsc{Accuracy} & \textsc{AUC} & \textsc{Accuracy} & \textsc{AUC} \\ [0.5ex]
\midrule
\textsc{MILRNN*} & 80.62\% & 0.807 & 86.19\% & 0.910 \\ [1ex]
\textsc{AbMILP*} & 84.50\% & 0.865 & 77.19\% & 0.865 \\ [1ex]
\textsc{DSMIL*} & 86.82\% & 0.894 & 80.58\% & 0.892 \\ [1ex]
\textsc{CLAM-SB*} & 87.60\% & 0.881 & 81.80\% & 0.881 \\ [1ex]
\textsc{CLAM-MB*} & 83.72\% & 0.868 & 84.22\% & 0.937 \\ [1ex]
\textsc{TransMIL*} & \textit{88.37\%} & 0.931 & \textit{88.35\%} & \textit{0.960}\\ [1ex]
\textsc{ProtoMIL*} & 87.29\% & \textit{0.935} & 83.66\% & 0.918\\ [1ex]
\midrule
\textsc{instance+mean*} & 79.84\% & 0.762 & 72.82\% & 0.840 \\ [1ex]
\textsc{instance+max*} & \textbf{82.95\%} & \textbf{0.864} & 85.93\% & \textbf{0.946} \\ [1ex]
\textsc{ProMIL (our)} & \textbf{82.95\%} & \textbf{0.864} & \textbf{86.71\%} & 0.940\\ [1ex]
\bottomrule
\end{tabular}
\label{table:big_histo}
\end{table}

\begin{table}
\caption{Our \textsc{ProMIL} method achieves superior results for the Doppler ultrasound database. Other methods such as \textsc{instance+max}, and \textsc{instance+mean} obtained inferior results compared to \textsc{ProMIL}. Moreover, the performance of \textsc{AbMILP} for this task is reduced showing that our \textsc{ProMIL} which is an instance-based approach is more suitable for this task. By 'Accuracy' we mean balanced accuracy.}
\centering
\scriptsize 
\begin{tabular}{ccc}
\cmidrule{2-3}
& \multicolumn{2}{c}{\textsc{Doppler Ultrasound}} \\ [0.5ex]
\toprule
\textsc{Method} & \textsc{Accuracy} & \textsc{AUC}  \\ [0.5ex]
\midrule
\textsc{AbMILP} & $79.77\%$ & $0.85$ \\[1ex]
\midrule
\textsc{instance+max} & $70.09\%$ & $0.71$  \\ [1ex]
\textsc{instance+mean} & $75.67\%$ & $0.79$ \\ [1ex]
\textsc{ProMIL (our)} & $\textbf{86.36}\%$ & $\textbf{0.92}$\\[1ex]
\bottomrule
\end{tabular}
\label{tab:ecg}
\end{table}

\paragraph{MNIST-bag results.}
We conducted a comparative analysis of our ProMIL approach with the instance-max and instance-mean methods, as well as with Attention MIL Pooling~\cite{ilse2018attention}. As illustrated in~\Cref{fig:mnist_results}, our ProMIL outperforms all other methods in terms of AUC. Additionally, we found that percentage-based assumption is difficult for AbMILP and can lead to random performance on the test dataset, even if the model converges during training. Moreover, we observed that Instance-MAX and Instance-MEAN struggle to distinguish between positive and negative bags when the ground truth $q$ is in the range of $[0.1, 0.5]$. With an increase in bag size or the number of training bags, all models achieve better performance. However, our ProMIL still outperforms them, making it suitable for both small and large datasets with small or large bags.

\paragraph{Histopathological results.}
We evaluated the versatility of our ProMIL approach on three real-life datasets commonly used as standard benchmarks in MIL: Colon Cancer, Camelyon16, and TCGA-NSCLC. The results for Colon Cancer are presented in~\Cref{table:colon}, while those for Camelyon16 and TCGA-NSCLC are in~\Cref{table:big_histo}. Our ProMIL approach outperformed other Instance-based methods for Colon Cancer and TCGA-NSCLC and matched the performance of the Instance-MAX approach for Camelyon16. This similarity in performance can be attributed to the small amount of metastatic tissue required for a positive label in Camelyon16~\cite{camelyon16paper}, leading our ProMIL to learn a small $q$ what is making it similar to Instance-MAX. When compared to other MIL approaches that operate on the bag level, such as TransMIL and ProtoMIL, our ProMIL outperformed some of them (e.g., MILRNN for TCGA-NSCLC and AbMILP for Colon Cancer) and was slightly worse than recent approaches using transformer-based architectures such as TransMIL. These results demonstrate the adaptability of ProMIL to a variety of MIL problems in the digital pathology field.

\paragraph{Doppler Ultrasound database results.}
We conducted an evaluation of our approach on a dataset of ultrasound videos with varying numbers of frames. Similar to the previous experiment, we compared our method with the baseline (instance-based) approaches. The results of this evaluation are presented in~\Cref{tab:ecg}.

\our{} utilizes a unique strategy for selecting the most significant frames in each video using the $q=38.56\%$-quantile criterion. This approach contrasts with the competing methods, \textsc{instance+max} and \textsc{instance+mean}, which rely on single-frame and all-frame analyses, respectively. In our evaluation, we found that choosing the dominant prediction did not necessarily lead to high-performance scores. Furthermore, using all frames, as done in the \textsc{instance+mean} method, produced better results than the \textsc{instance+max} method, but still failed to achieve significantly better results, indicating overfitting to the noise.

Our approach's superiority over the baseline methods is further highlighted by the effectiveness of selecting the most significant frames in each video, allowing us to mitigate the effects of noise and achieve superior performance.

\begin{figure}[t]
    \centering
    \includegraphics[width=0.97\columnwidth]{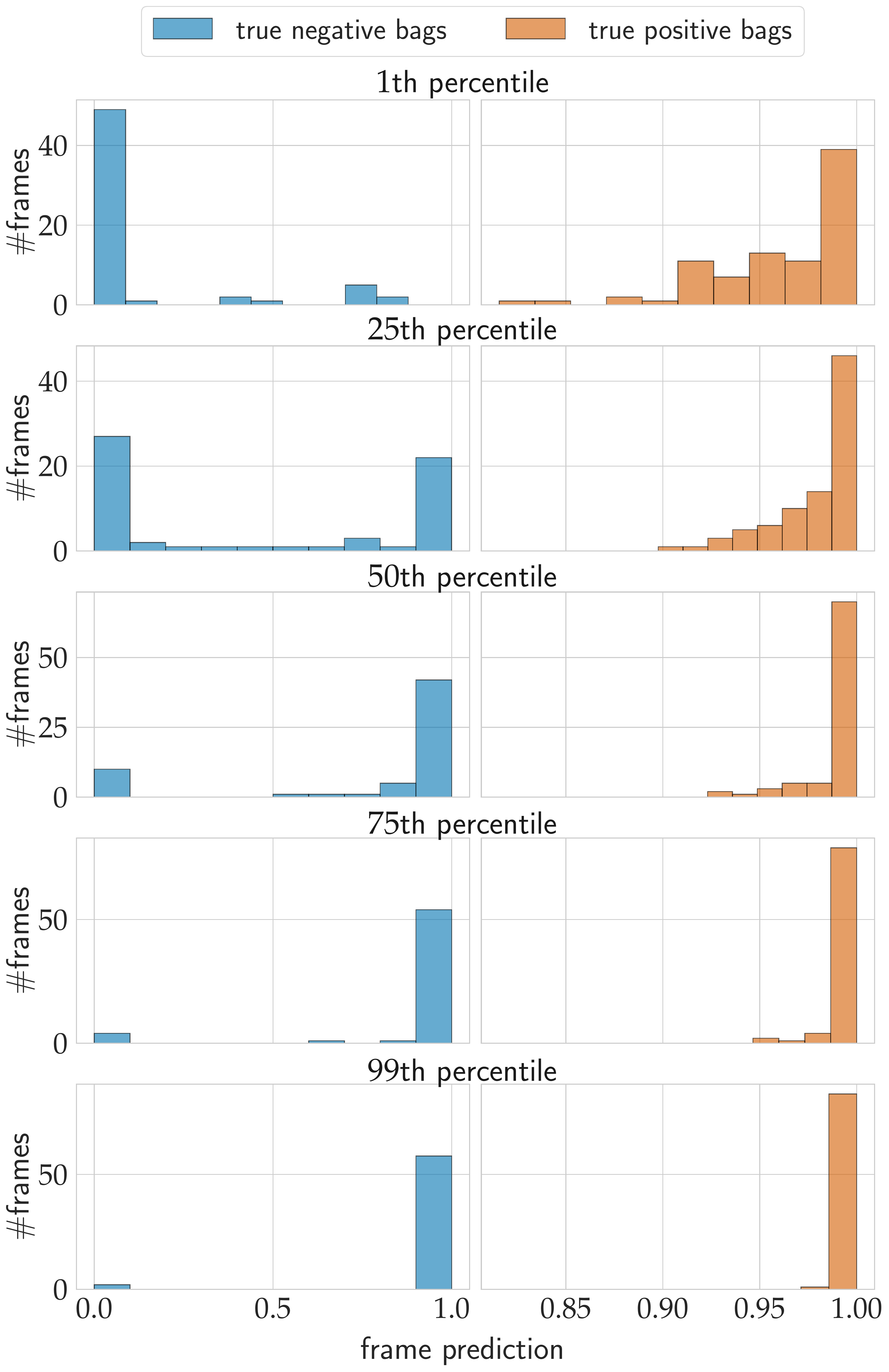}
    \caption{The illustration depicts the classification predictions generated by the ProMIL $f$ classification component. This component relies on the selection of significant frames, representing 38.56\% of the frames with the highest prediction values, that impacted its decision-making process. The figure displays the 1st, 25th, 50th, 75th, and 99th percentiles of these significant frames in each row. The left column exhibits the prediction distribution for true negative bags, while the right column shows the distribution for true positive bags. Note that frames with predictions below threshold 0.5 are absent from the right column, which is not apparent from the left column. Even among the 99th percentile, there are frames with predictions approaching zero.}
    \label{fig:ProMIL_hist}
\end{figure}

Furthermore, \Cref{fig:ProMIL_hist} provides insights into the distribution of the predictions made by the $f$ classification component of the ProMIL model. This distribution is based on the selection of significant frames, which our model determined to be 38.56\% for X data. To illustrate this, we only considered true negative and positive bags and presented their prediction distributions separately. Our results show that for true positive bags, all significant frames have predictions above 0.8, indicating high confidence in their classification. Conversely, for true negative bags, up to the 50th percentile, instances with predictions close to zero dominate the model's inference.

In addition to analyzing the number and selection of frames, our study also focused on identifying the specific objects and regions of interest in individual frames that our model considers when making predictions. To achieve this, we utilized the widely recognized Grad-CAM technique~\cite{selvaraju2017grad}. This approach generates a heatmap highlighting the critical regions of an input image that the model focused on to arrive at its decision.

\begin{figure}[t]
  \centering
  \includegraphics[width=0.97\columnwidth]{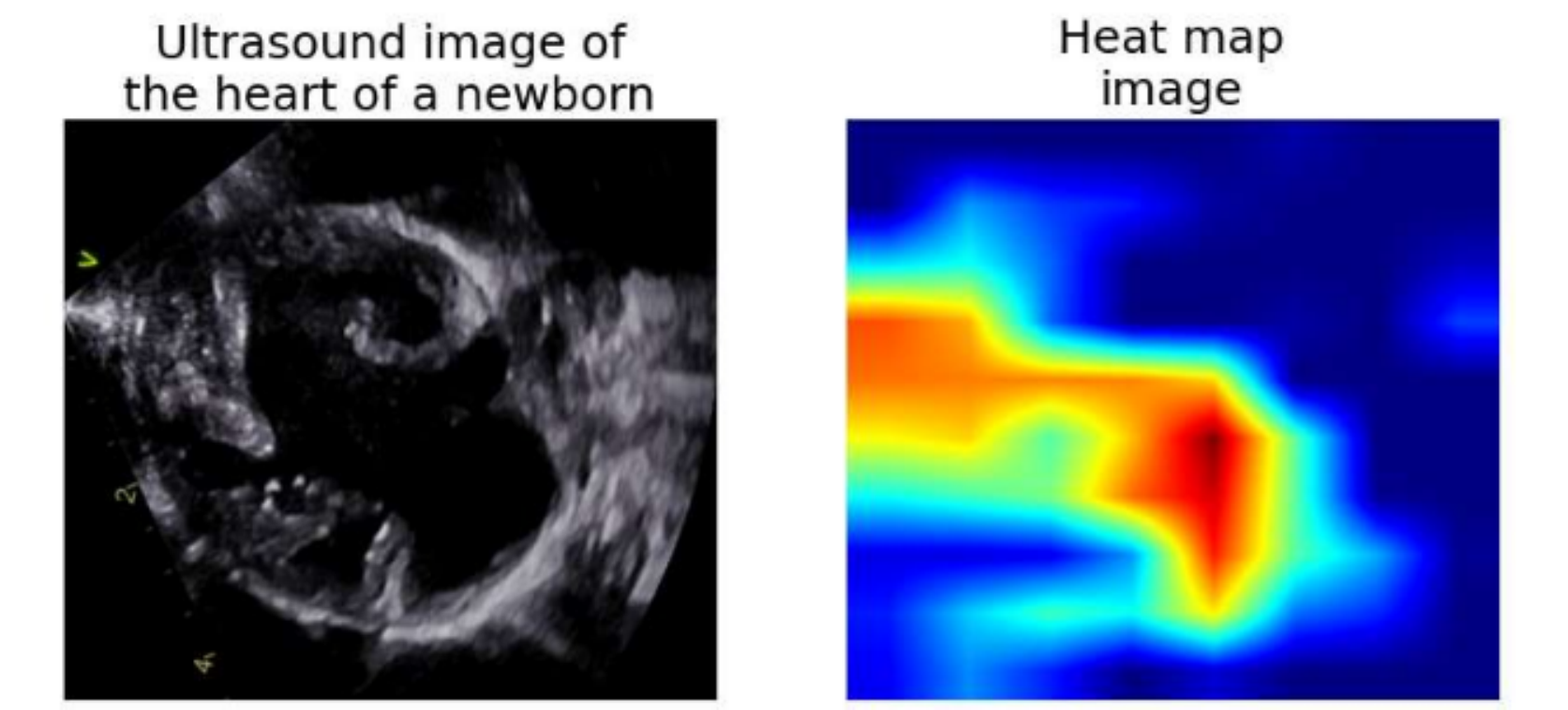}
  \caption{An ultrasonographic image of a newborn's heart with a ventricular septal defect and the corresponding heat map. This map allows to understand which features of the image were important for the model during the prediction process, taking into account the differences between successive frames of the processed and analyzed ultrasound recording. Areas with larger changes on the heat map are marked in yellow, red and burgundy as areas with higher heat intensity. Consultations with medical experts in the field of cardiology confirmed that such areas in the presented image indicate the locations of the ventricular septal defect.}
  \label{fig:heatmap_ultrasond}
\end{figure}

\Cref{fig:heatmap_ultrasond} presents several results that demonstrate the heart areas our model emphasized when examining ultrasound images of a child with a congenital heart defect that involves a septal defect. After consulting with clinical experts in the field of pediatric cardiology, we confirmed that the regions marked on the heatmaps align with the areas typically identified by experts. Consequently, our approach could assist medical practitioners in comprehending the outcomes and recognizing heart defects in newborns.

\section{Conclusions}

In this paper, we propose a novel instance-based approach \our{} to Multiple Instance Learning (MIL), where bag classification is obtained by aggregating instance-level predictions. It solves the percentage-based assumption, which occurs in many real-world medical applications, using Bernstein polynomial estimation.

\our{} outperforms other instance-based approaches and obtains results comparable to less interpretable representation-based models. Therefore,  it can be easily applied to critical domains, such as computer-assisted interventions.

\section{Acknowledgements}

This research was partially funded by the National Science Centre, Poland, grants no.  2020/39/D/ST6/01332 (work by \L{}ukasz Struski), 2022/47/B/ST6/03397 (work by Bartosz Zieli\'nski), and 2022/45/N/ST6/04147 (work by Dawid Rymarczyk). Moreover, Dawid Rymarczyk received an incentive scholarship from the funds of the program Excellence Initiative -- Research University at the Jagiellonian University in Kraków. Works of Jacek Tabor, Arkadiusz Lewicki, and Robert Sabiniewicz were supported by company the UES Ltd. 
Some experiments were performed on servers purchased with funds from a grant from the Priority Research Area (Artificial Intelligence Computing Center Core Facility) under the Strategic Programme Excellence Initiative at Jagiellonian University.

\bibliography{ref}
\end{document}